\pdfoutput=1
\documentclass[sigconf,nonacm]{acmart}
\makeatletter\let\hyxmp@parse@acmart\relax\makeatother
\usepackage{pifont}
\usepackage{array}
\usepackage{tabularx}
\usepackage{float}
\AtBeginDocument{%
  \renewcommand{\checkmark}{\ding{51}}}

\acmConference[MODELS '26]{ACM/IEEE International Conference on Model Driven Engineering Languages and Systems}{October 4--9, 2026}{M\'alaga, Spain}
\acmYear{2026}
\copyrightyear{2026}

\AtBeginDocument{\let\balance\relax}

\begin{document}

\title{Live Architecture Models for Cloud-Native Architecture-as-Code: Early Results from Kubernetes Conformance Checking}

\settopmatter{authorsperrow=2}
\author{Denis Mamatin}
\affiliation{%
  \department{PhD student}
  \institution{Moscow Institute of Physics and Technology (MIPT)}
  \country{}
}

\author{Andrey Salov}
\affiliation{%
  \department{PhD}
  \institution{Kangwon National University}
  \country{}
}

\begin{abstract}
Architectural consistency is critical in software engineering, directly affecting technical debt, maintainability, and system reliability. A central challenge in software evolution is architecture erosion (AEr), where the implemented system diverges from its intended architecture. In cloud-native systems, one observable source of such divergence is infrastructure drift: the deployed Kubernetes state no longer matches the architectural intent documented by the team. The approaches reviewed here address architecture conformance, runtime models, and deployment governance, but do not combine editable KDL architectural intent with recurring Kubernetes conformance diagnostics in an IDE.
This paper investigates live architecture models, a modeling approach in which architectural representations are connected to a running system through explicit correspondence links and recurring conformance checks. The claim investigated is that such models can make cloud-native Architecture-as-Code runtime-informed: architectural intent remains editable in the modeling environment, while selected Kubernetes runtime state can be recovered and compared against that intent. The approach is instantiated through KDL, a domain-specific language for Kubernetes deployment architecture, and Archer, a VS Code prototype that provides view synchronization between textual and graphical architectural views. Within the current KDL scope, Archer recovers selected architectural elements from the Kubernetes API and compares the recovered runtime-informed model with user-authored KDL specifications during periodic and on-demand checks.
We assess feasibility on three feature-selected Kubernetes example applications under an author-defined protocol, measuring precision and recall for snapshot recovery and selected inconsistency detection. The results support feasibility within the evaluated KDL scope, without establishing comparative superiority or detection of arbitrary production drift.

\end{abstract}

\begin{CCSXML}
<ccs2012>
   <concept>
       <concept_id>10011007.10011006.10011060.10011063</concept_id>
       <concept_desc>Software and its engineering~System modeling languages</concept_desc>
       <concept_significance>300</concept_significance>
       </concept>
   <concept>
       <concept_id>10011007.10011006.10011073</concept_id>
       <concept_desc>Software and its engineering~Software maintenance tools</concept_desc>
       <concept_significance>300</concept_significance>
       </concept>
   <concept>
       <concept_id>10011007.10011006.10011060.10011062</concept_id>
       <concept_desc>Software and its engineering~Architecture description languages</concept_desc>
       <concept_significance>100</concept_significance>
       </concept>
 </ccs2012>
\end{CCSXML}

\ccsdesc[300]{Software and its engineering~System modeling languages}
\ccsdesc[300]{Software and its engineering~Software maintenance tools}
\ccsdesc[100]{Software and its engineering~Architecture description languages}

\keywords{Architecture as Code, Architecture Erosion, View Synchronization, Infrastructure Drift, Model-Driven Engineering, Kubernetes Architecture, Domain-Specific Language (DSL)}

\maketitle

\section{Introduction}

Modern software systems continuously face the need for architectural changes driven by increasing demands for scalability and distribution. These changes often lead to divergence between the intended software architecture and its actual implementation, resulting in phenomena such as Architecture Erosion (AEr) \citep{liUnderstandingSoftwareArchitecture2022}. In this paper, architecture erosion denotes the long-term degradation of the intended architecture as implementation and operational decisions evolve away from it. AEr manifests as the gradual loss of the system's original structure, increased maintenance complexity, and the accumulation of technical debt. One contributing factor is weak coupling between the architectural model and the actual implementation, particularly in environments that rely heavily on manual artifact alignment and lack sufficient automation for conformance checking.

Integration of architecture description languages (ADLs) and modeling tools with development environments and automated conformance checking remains a challenge \citep{andrewsSoftwareArchitectureErosion2020}. As a result, there is growing interest in the concept of AaC, which involves using formal, machine-readable architectural descriptions stored alongside source code and kept aligned with it. AaC aims to support alignment, traceability, and change management. In this paper, Architecture as Code (AaC) denotes the general paradigm, live architecture models denote our modeling realization of that paradigm, and Archer denotes the concrete VS Code-based prototype that implements it.

One operational contributor to architecture erosion in cloud-native systems is infrastructure drift. Infrastructure drift denotes a mismatch between desired deployment configuration or architecture model and live deployed infrastructure state. Such divergence complicates infrastructure management, reduces system reliability, and may lead to operational failures. If left unchecked, infrastructure drift can escalate into AEr, where infrastructure-level changes become ad hoc and uncoordinated, ultimately degrading system maintainability and hindering future evolution \citep{Bucaioni7118}.

The absence of automated conformance checking between the architectural model and the actual system allows their gradual divergence to remain unnoticed. This inconsistency complicates system analysis, hinders the onboarding of new developers, and increases the risk of maintenance errors \cite{desilvaControllingSoftwareArchitecture2012}. While architectural modeling tools and domain-specific languages (DSLs) do exist, they are often isolated from the development workflow and unaware of the system's actual configuration. There is a clear need for a solution that not only supports defining software AaC but also performs automated conformance checking against the current state of the deployment environment. In addition, view synchronization between graphical and textual representations of software architecture remains a pressing challenge \cite{davidBlendedModelingCommercial2022}. Without effective view synchronization, users must manually maintain alignment between textual and graphical representations, which can reduce the efficiency and reliability of architecture analysis tools.

Here, architectural intent means the selected deployment structure and properties authored in KDL, not high-level goals automatically refined into deployment actions.

This paper uses the following operational terms. Recovery denotes snapshot extraction of selected deployment facts from the Kubernetes API into KDL elements. Conformance checking denotes diagnostic comparison between architectural intent and recovered or live implementation facts. Detection denotes reporting an inconsistency after it exists. Prevention denotes blocking, enforcing, or repairing a violating change before it persists; Archer does not implement prevention. View synchronization denotes keeping the textual KDL document and graphical diagram aligned during editing. A live architecture model denotes an editable architecture model linked to selected runtime state through correspondence links and recurring conformance checks. Here, recurring refers to read-only checks; snapshot recovery is a separate, explicitly invoked operation and does not continuously merge runtime changes into the authored model.

These definitions separate three lifecycle concerns. Design time denotes authoring and editing architectural intent as KDL and synchronized diagrams inside the IDE. Runtime observation denotes querying a running Kubernetes cluster and recovering selected deployment facts through the Kubernetes API. Archer does not turn runtime activity into a design-time mechanism; it brings runtime feedback into a design-time modeling environment. Deployment-time reconciliation and enforcement, such as applying manifests, blocking invalid resources, or reconciling a cluster with Git, are outside the current prototype scope.

The central thesis of this paper is that live architecture models can make cloud-native AaC runtime-informed by connecting architectural intent, expressed as an editable model, with selected deployment facts recovered from a running Kubernetes cluster. The prototype targets detection rather than direct erosion control. It addresses one prerequisite for controlling erosion: making divergences between architectural intent and deployed Kubernetes state observable inside the modeling workflow.

The contribution concerns editable architectural intent and its Kubernetes-specific realization, not a new general conformance-checking algorithm.

The paper makes four contributions spanning the concept, its technical realization, and early empirical evidence. First, it defines live architecture models for cloud-native AaC as architectural models connected to selected runtime deployment facts through explicit correspondence and recurring conformance checks. Second, it instantiates this idea for Kubernetes deployment architecture through KDL and deterministic recovery and comparison rules within a stated coverage boundary. Third, it realizes the approach in Archer, a VS Code prototype that keeps textual and graphical KDL views synchronized and reports model-cluster inconsistencies. Fourth, it provides feasibility evidence from three feature-selected Kubernetes example applications under an author-defined protocol, measuring recovery and inconsistency-detection precision and recall within the supported KDL scope.

\section{Related Work}

Research on AEr has produced techniques for architecture conformance checking, consistency preservation, runtime models, architecture recovery, blended modeling, and deployment governance. This section positions Archer against five related streams: architecture conformance checking and consistency preservation; runtime models and digital architecture twins; blended textual/graphical modeling and diagramming; modeling and architecture recovery; and Infrastructure as Code and Kubernetes deployment governance. The comparison focuses on the capabilities relevant to Archer: textual or model support, diagram support, external system facts used for checking or recovery, and conformance-checking or governance support.

\subsection{Architecture Conformance Checking}

Architecture conformance checking studies whether an implementation conforms to an intended architectural description. A typical conformance workflow distinguishes three artifacts: an intended architecture, a set of implementation facts, and a mapping between them. The check then classifies relationships between expected and observed facts, commonly identifying convergences, divergences, and absences.

Software Reflexion Models are the foundational example of this mapping-based view of conformance checking \citep{murphySoftwareReflexionModels2001}. In that approach, a high-level model is compared with source-code facts through a developer-specified mapping, making architectural drift observable without requiring the implementation to be written in a specialized architecture language. ReflexML \citep{adersbergerReflexMLUMLBasedArchitecturetoCode2011} follows the same broad problem setting in a UML-based context by attaching mappings from architecture models to code elements and applying predefined consistency checks. ArchJava \citep{aldrichArchJavaConnecting2002} addresses a related problem from the programming-language side by embedding architectural structure into Java so that implementation code can be checked against architectural constraints.

Beyond detection-oriented conformance checking, change-driven consistency approaches aim to keep related artifacts aligned as they evolve. \citet{kramerChangeDrivenConsistency2015} present an approach for maintaining consistency among component code, architectural models, and contracts by reacting to changes in one artifact and propagating or checking their effects on the others. This line of work is closer to consistency preservation than to diagnostic conformance checking because consistency is treated as an ongoing multi-artifact maintenance problem.

Archer builds on this conformance-checking lineage rather than replacing it. Its distinction is the target and workflow: the compared artifacts are not primarily source-code facts and architecture diagrams, but user-authored KDL deployment architecture and selected runtime facts recovered from a Kubernetes cluster. Archer also embeds the check into an AaC modeling workflow by keeping textual and graphical views synchronized and reporting model-cluster inconsistencies in the IDE. Unlike change-driven consistency-preservation approaches, the current prototype does not propagate changes across artifacts and does not enforce, block, or repair violating changes. It detects mismatches and makes them visible in the architectural modeling workflow.

\subsection{Runtime Models and Digital Architecture Twins}

Models@Runtime studies the use of software models during system execution, where models provide an abstraction for monitoring, analysis, reasoning, and adaptation of running systems \citep{bencomoUseSoftwareModels2009}. This literature is closely related to Archer because Archer also connects an architectural model with runtime information obtained from the running system. The relation is conceptual: in both cases, a model is not merely documentation but an artifact that can be compared with or informed by execution-time facts.

To address architectural drift and support software maintenance tasks, \citet{jordanAutoArxDigitalTwins2022} propose the Digital Architecture Twin (DArT), a dynamic architectural data model that evolves alongside the underlying software system. DArT enables continuous recovery and aggregation of architectural information from various sources, including source code, build and deployment scripts, and architectural views. This model facilitates integration into the development process through a continuous reverse engineering mechanism, thereby providing stakeholders with up-to-date and context-relevant architectural information.
However, \citet{ammermannQueryLanguageSoftware2023} highlight the lack of effective interaction mechanisms between users and the DArT model. To address this limitation, they introduce Architecture Information Query Language (AIQL), a dedicated query language designed for stakeholders ranging from developers to solution architects. Their user study reports that the language helps users retrieve relevant architectural information and can scale to large software systems.

Archer differs from this line of work in scope and intent. It does not provide a general runtime model for analysis or adaptation, and it does not autonomously evolve the running system. Instead, it focuses on cloud-native deployment architecture as an AaC artifact: KDL remains editable in the IDE, selected Kubernetes deployment facts are recovered from the running cluster, and conformance diagnostics expose differences between the model and the cluster state. Thus, Archer uses runtime information to support architectural conformance checking and modeling feedback rather than runtime adaptation.

\subsection{Blended Modeling and Diagramming}

Several recent studies address blended textual and graphical modeling for DSLs and modeling tools. For instance, \citet{Glaser_Bork_2021} present the bigER tool, which combines textual and graphical modeling of Entity Relation Diagrams (ERD) using Xtext \citep{eclipseXtextLanguage} and Sprotty \citep{eclipseEclipseSprotty}. The tool integrates with VS Code, supports interactive diagrams, Structured Query Language (SQL) code generation, and model validation. Its scope is data modeling, so it does not address architecture-level conformance against implementation or runtime system state. The work by \citet{metinReferenceArchitectureDevelopment2025} focuses on the development of an extensible architecture for web-based modeling tools using the Graphical Language Server Platform (GLSP) \citep{eclipseGLSP}. This architecture underpins the bigUML editor, which provides graphical Unified Modeling Language (UML) modeling capabilities and integration with VS Code. The work is relevant as a modeling-tool architecture, but it does not address architecture conformance against the actual system state. A closely related implementation pattern is presented by \citet{lencsesCombiningTextualGraphical2024}, where graphical and textual modeling are integrated through a shared model using GLSP, Langium \citep{langiumLangium}, and a Model Server. This approach supports synchronized updates, extensibility via dependency injection, and integration with modern IDEs.

These works show that synchronized textual and graphical modeling is an established blended-modeling pattern rather than a standalone contribution of Archer. In Archer, view synchronization is enabling infrastructure: it makes KDL usable as an AaC artifact in an IDE while keeping the textual and graphical representations aligned. The distinguishing element is the combination of this blended modeling workflow with Kubernetes-specific architectural recovery and recurring conformance diagnostics against selected runtime deployment facts.

Traditional tools for creating architectural diagrams typically follow the boxes and arrows paradigm. These tools are useful for communication, but they are usually separated from version-controlled architecture artifacts and automated conformance checks. The Diagrams as Code (DaC) approach addresses part of this limitation by representing diagrams in code form. This enables team collaboration, versioning, and automated visualization generation. Some of the most widely adopted tools in this space include PlantUML \citep{plantumlOpensourceTool}, which generates UML diagrams from textual descriptions; Mermaid \citep{mermaidMermaid}, which uses Markdown-inspired text definitions for diagrams and charts; and Structurizr \citep{structurizrStructurizr}, which supports model-as-code architecture descriptions for the C4 model. These tools support code-centered diagram production, but they are not designed to recover live deployment facts or check an architecture model against a running Kubernetes system.
A visual modeling approach is exemplified by IcePanel \citep{icepanelIcePanelCollaborative}, a collaborative software architecture modeling tool based on the C4 model. IcePanel supports hierarchical C4 views and reusable model elements. Its focus is visual architecture modeling and documentation rather than IDE-based DSL editing, static language services for an architecture DSL, or runtime-informed conformance checking against a deployed system.

\subsection{Modeling and Architecture Recovery}

\citet{haitzerDSLbasedSupportSemiautomated2012} propose a semi-automated strategy to support traceability between architectural models and source code. The goal of their method is to provide architectural abstraction specifications using a DSL. The approach extracts code-level information and uses DSL-defined abstraction specifications to generate UML component and connector views. It also supports traceability links between architectural elements and source code artifacts, comparison of generated architectural views, and consistency checking of architectural design constraints. This work is directly relevant because it uses a DSL to connect architecture abstractions with implementation facts. It differs from Archer in its target and data sources: it focuses on source-code-based architectural abstraction, while Archer focuses on Kubernetes deployment architecture, KDL models, and selected runtime facts recovered through the Kubernetes API.

For the purpose of automated software architecture verification, an industry report by \citet{GovernanceAsCode2024} describes the Governance as Code approach. To describe software architecture, the authors utilize a digital architecture data format, which represents a collection of architectural features. These features are first registered in a centralized feature registry, from which the software architecture description is constructed.
Developers can interact with the digital architecture data format both at design time and during release preparation before production deployment. In the latter case, the digital architecture must first be recovered from various sources, including source code, configuration files, system metrics, and other relevant artifacts. Once the digital software architecture format is available, the verification process can be automated. All architectural rules are encoded as small programs that are applied to the digital architecture model. This results in a toolchain that automatically checks a system's architecture against a set of rules defined by the development team.

\subsection{Infrastructure as Code}

GitOps operationalizes IaC practices for Kubernetes by storing desired deployment state in version-controlled repositories and continuously comparing that state with the live cluster. Argo CD \citep{argoCDDocs} is a declarative GitOps continuous delivery tool implemented as a Kubernetes controller that monitors running applications and compares live state with the desired target state stored in Git. Flux \citep{fluxDocs} similarly keeps Kubernetes clusters reconciled with configuration sources such as Git repositories and supports reconciliation of Kubernetes resources, including Helm and Kustomize workflows.

These tools are directly relevant because they address drift in practical Kubernetes delivery pipelines. However, their primary source of truth is the operational deployment configuration: manifests, Helm charts, Kustomize overlays, or other configuration sources. Archer works at a different abstraction level. Its source of architectural intent is a KDL model edited in an IDE and synchronized with a graphical view. GitOps tools reconcile cluster state with desired deployment artifacts; Archer compares architectural deployment intent with selected runtime facts and reports model-cluster inconsistencies. The approaches are complementary: GitOps can keep a cluster aligned with deployment artifacts, while Archer can provide architecture-level modeling and conformance diagnostics over Kubernetes deployment structures.

Policy-as-code tools provide another practical mechanism for checking Kubernetes resources. OPA Gatekeeper \citep{opaGatekeeperDocs} integrates OPA with Kubernetes admission control through constraints, constraint templates, and audit functionality. Kyverno \citep{kyvernoDocs} provides a Kubernetes-native policy engine with policy types for validating, mutating, generating, and deleting resources, as well as validating container images. These tools can check resources before or during admission to the cluster and can therefore block policy-violating resources.

Cloud-native application and platform modeling frameworks raise the abstraction above raw manifests. Open Application Model (OAM) defines higher-level abstractions for cloud-native applications and represents applications through components, traits, scopes, and application objects \citep{oamSpec}. KubeVela uses OAM as its application delivery model: an application deployment plan is composed from components, traits, policies, and workflow steps, and KubeVela uses Kubernetes APIs and control loops to drive the underlying infrastructure capabilities \citep{kubeVelaApplicationDocs}. Crossplane takes a platform-control-plane view: composite resource definitions define custom APIs, compositions specify how those APIs create composed resources, and managed resources map Kubernetes objects to external provider resources whose \texttt{forProvider} fields are treated as the source of truth \citep{crossplaneCompositionsDocs,crossplaneManagedResourcesDocs}. CAMEL similarly targets cloud application modeling, but in a multi-cloud setting: it specifies design-time aspects of cloud applications, supports models@runtime for current-state capture and adaptive provisioning, and aligns with TOSCA \citep{achilleosCloudApplicationModelling2019}.

OAM/KubeVela, Crossplane, and CAMEL use their models to drive deployment, provisioning, or adaptation. Archer uses its model for IDE-based authoring and diagnostic comparison.

Archer addresses a different model-cluster comparison problem. Gatekeeper and Kyverno express expected properties as policies over Kubernetes resources; Archer expresses architectural intent as a KDL deployment model and compares it with selected runtime deployment facts. Policy-as-code tools are stronger for admission control and resource-level governance. Archer is intended for architecture-level modeling feedback: it shows the modeled deployment structure, keeps textual and graphical views aligned, and reports differences between the architectural model and the live cluster state. A future workflow could combine these directions, for example by deriving policy checks from KDL or by running Archer conformance checks before GitOps reconciliation.

\citet{soldaniOfflineMiningMicroserviceBased2023} proposed an offline methodology, implemented in $\mu$TOM, for the automatic recovery of the architecture of microservice-based applications deployed in Kubernetes environments. The approach consists of two main phases. First, a draft architectural model is constructed in the $\mu$TOSCA format using Kubernetes manifests and the interaction graph generated by Kiali—an observability tool based on Istio. This initial model is then refined by incorporating component types (e.g., services, databases, message brokers) and interaction characteristics such as timeouts, circuit breakers, and dynamic service discovery. The resulting architecture can be visualized and analyzed using tools from the $\mu$TOSCA ecosystem, such as microFreshener.
However, the approach has several limitations. First, the correctness of component classification relies on the use of official Docker images: if a component is implemented using a custom image, it may be incorrectly classified as a generic service, even if it actually functions as a database or message broker. Second, it is an offline mining technique: it produces a $\mu$TOSCA representation from Kubernetes manifests and a Kiali graph, but it does not maintain an editable AaC model aligned with source code or cluster state inside an IDE.

In a master's thesis, \citet{figueiredo2024recovery} recovers C4 level-2 container diagrams from Terraform scripts and renders them in PlantUML. The approach uses static IaC analysis; Archer instead obtains selected deployment facts from the live Kubernetes API and compares them with an editable KDL model.

\begin{table*}[t]
\caption{Qualitative comparison of the discussed approaches}
\label{tab:related-works}
\centering
\small
\renewcommand{\arraystretch}{1.05}
\setlength{\extrarowheight}{0pt}
\setlength{\tabcolsep}{6pt}
\begin{tabular}{@{}lcc lc@{}}
\toprule
\textbf{Approach} & \textbf{Text/model} & \textbf{Diagram} & \textbf{External facts} & \textbf{Checking} \\
\midrule
\multicolumn{5}{l}{\itshape Conformance checking} \\
Reflexion Models \cite{murphySoftwareReflexionModels2001} & \texttimes & \texttimes & Source code & \checkmark \\
ReflexML \cite{adersbergerReflexMLUMLBasedArchitecturetoCode2011} & \checkmark & Generation & Source code & \checkmark \\
ArchJava \cite{aldrichArchJavaConnecting2002} & \checkmark & \texttimes & Source code & \checkmark \\
Change-driven consistency \cite{kramerChangeDrivenConsistency2015} & \texttimes & \texttimes & Code/models & \checkmark \\
\midrule
\multicolumn{5}{l}{\itshape Modeling, diagramming, and recovery} \\
bigER \cite{Glaser_Bork_2021} & \checkmark & \checkmark & \texttimes & \checkmark \\
bigUML \cite{metinReferenceArchitectureDevelopment2025} & \texttimes & \checkmark & \texttimes & \texttimes \\
Langium + GLSP \cite{lencsesCombiningTextualGraphical2024} & \checkmark & \checkmark & \texttimes & \texttimes \\
DaC tools \cite{plantumlOpensourceTool,mermaidMermaid,structurizrStructurizr} & \checkmark & Generation & \texttimes & \texttimes \\
Haitzer et al. \cite{haitzerDSLbasedSupportSemiautomated2012} & \checkmark & Generation & Source code & \checkmark \\
Governance as Code \cite{GovernanceAsCode2024} & \checkmark & \checkmark & Source/config & \checkmark \\
\midrule
\multicolumn{5}{l}{\itshape Runtime models} \\
DArT \cite{jordanAutoArxDigitalTwins2022} & \texttimes & Generation & Runtime/artifacts & \texttimes \\
\midrule
\multicolumn{5}{l}{\itshape Infrastructure as Code} \\
Argo CD \cite{argoCDDocs} & \texttimes & \texttimes & Live cluster & \checkmark \\
Flux \cite{fluxDocs} & \texttimes & \texttimes & Live cluster & \checkmark \\
OPA Gatekeeper \cite{opaGatekeeperDocs} & \texttimes & \texttimes & Live cluster & \checkmark \\
Kyverno \cite{kyvernoDocs} & \texttimes & \texttimes & Live cluster & \checkmark \\
OAM/KubeVela \cite{oamSpec,kubeVelaApplicationDocs} & \checkmark & \texttimes & Kubernetes API & \checkmark \\
Crossplane \cite{crossplaneCompositionsDocs,crossplaneManagedResourcesDocs} & \checkmark & \texttimes & Provider/Kubernetes & \checkmark \\
CAMEL \cite{achilleosCloudApplicationModelling2019} & \checkmark & \texttimes & Runtime/cloud & \texttimes \\
$\mu$TOM \cite{soldaniOfflineMiningMicroserviceBased2023} & \texttimes & Generation & Manifests/Kiali & \texttimes \\
Recovery IaC \cite{figueiredo2024recovery} & \texttimes & Generation & Terraform files & \texttimes \\
\midrule
\textbf{Archer} & \checkmark & \checkmark & Kubernetes API & \checkmark \\
\bottomrule
\end{tabular}
\par\smallskip
\begin{minipage}{0.96\textwidth}
\footnotesize
Author-interpreted categories, not a comparative accuracy measure. Checking includes diagnostics, reconciliation, and admission. A cross indicates no support identified in the cited discussion for that category, not absence throughout an ecosystem; ``Generation'' denotes diagram generation.
\end{minipage}
\end{table*}

Table~\ref{tab:related-works} summarizes this qualitative positioning. Archer combines editable KDL models, synchronized views, and Kubernetes recovery and diagnostics within a stated scope. The table distinguishes workflows and does not establish comparative accuracy or superiority.

\section{Architecture as code}

\subsection{Definition}

\citet{Bucaioni7118} define AaC as an approach in which software architecture is continuously defined, managed, and evolved through a machine-readable, version-controlled code base. We focus on two aspects: code-centric architectural artifacts and automation that connects them to runtime feedback. These aspects motivate editable KDL models, synchronized diagrams, recovery, and conformance diagnostics; they do not constitute an exhaustive operationalization of AaC.

\subsection{Kubernetes Deployment Language}

\begin{figure}[htbp]
    \centering
    \includegraphics[width=\linewidth]{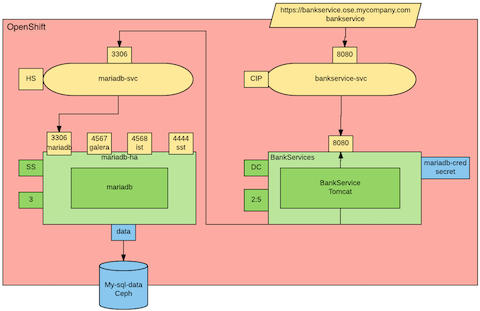}
    \caption{An example of Kubernetes cluster architecture described in KDL notation}
    \Description{A Kubernetes architecture diagram in KDL notation with nodes and links between core resources.}
    \label{fig:kdl-notation-example}
\end{figure}

KDL was introduced as a compact graphical notation for Kubernetes application deployments in an industry blog post \cite{KDLNotationDescribe}. In this paper, KDL denotes Archer's implemented subset of that notation, with a Langium grammar and the metamodel described in Section~\ref{sec:modeling-language}. The original notation and Archer's formalized subset are distinct. The notation illustrates architecturally significant deployment elements (Fig.~\ref{fig:kdl-notation-example}); Archer supports synchronized textual and graphical editing of its implemented subset.

\subsection{Conceptual model}

\begin{figure*}[t]
    \centering
    \includegraphics[width=\textwidth]{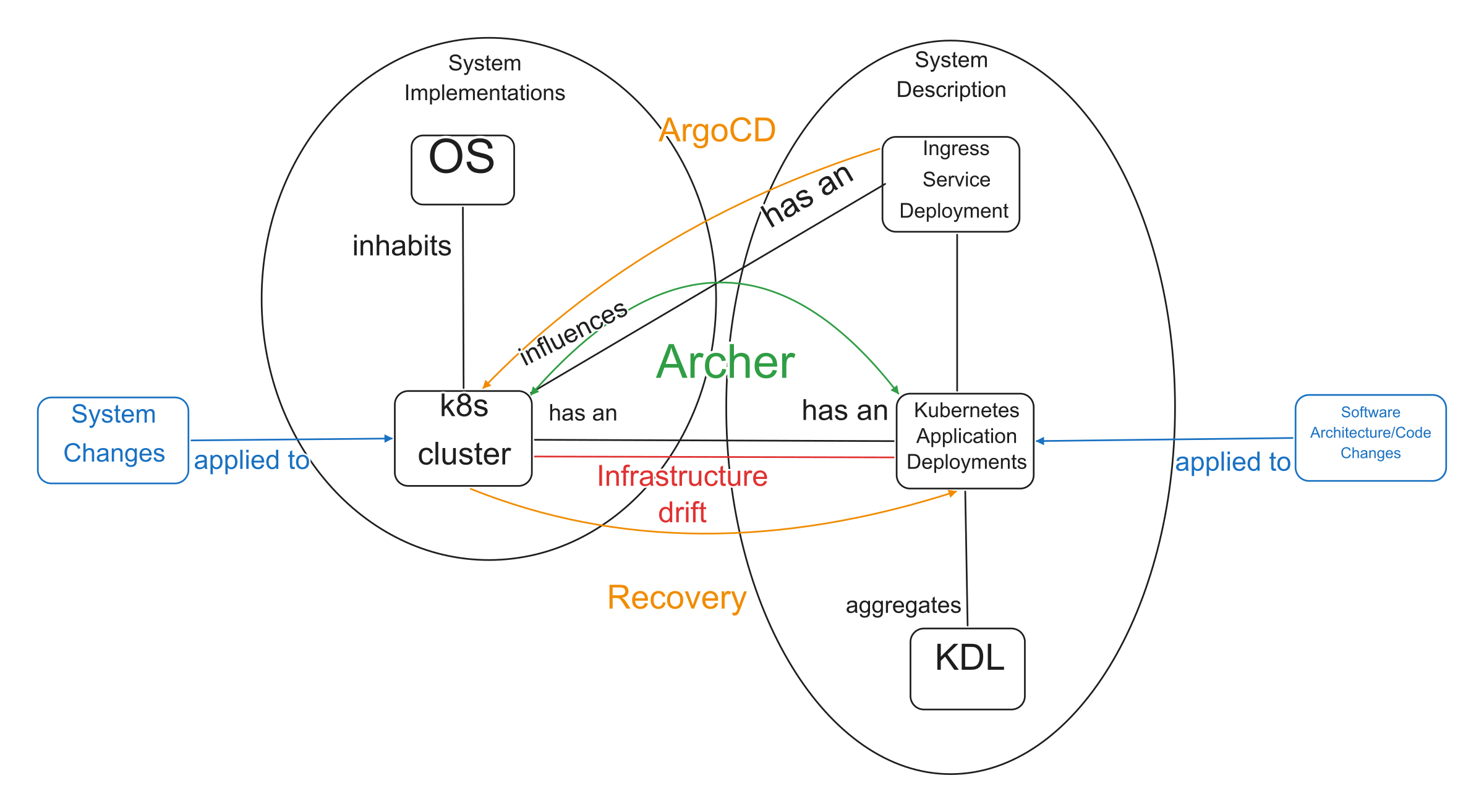}
    \caption{Conceptual overview of the Archer live architecture model workflow. The model relates runtime Kubernetes state and architectural description, contrasts drift and Argo CD-style reconciliation with Archer's recovery and conformance-checking feedback, and shows where runtime feedback enters the architecture-modeling workflow.}
    \Description{A conceptual model relating runtime Kubernetes state and architectural description, showing infrastructure drift, Argo CD-style reconciliation, and Archer recovery and conformance-checking feedback.}
    \label{fig:concept-model}
\end{figure*}

To implement the proposed approach, a conceptual model of architectural description was developed (see Fig.~\ref{fig:concept-model}), informed by the architectural-description concepts of IEEE Std 1471-2000 \citep{arc-1471-2000}. The contemporary standard is ISO/IEC/IEEE 42010:2022 \citep{iso42010_2022}; the mapping below is an illustrative application of the earlier concepts, not a claim of compliance with every requirement of either edition. This model defines the key components of the architectural description, the relationships among them, and the correspondence between architectural artifacts and the actual system.

Figure~\ref{fig:concept-model} illustrates the infrastructure drift problem targeted by this work: configuration changes or Kubernetes cluster updates without corresponding updates to architectural models can create misalignment between architectural intent and the system's current state.

Figure~\ref{fig:concept-model} contrasts manifest-based deployment and GitOps reconciliation with Archer's feedback to an editable architecture model. Archer reports model-cluster mismatches through periodic and on-demand checks; Fig.~\ref{fig:kdl-architecture} shows the implementation.

The illustrative mapping from architecture-description concepts to Kubernetes artifacts is provided in Appendix~\ref{app:representation}.

\section{Archer}

\subsection{Archer architecture and workflow}
\begin{figure*}[t]
    \centering
    \includegraphics[width=\textwidth]{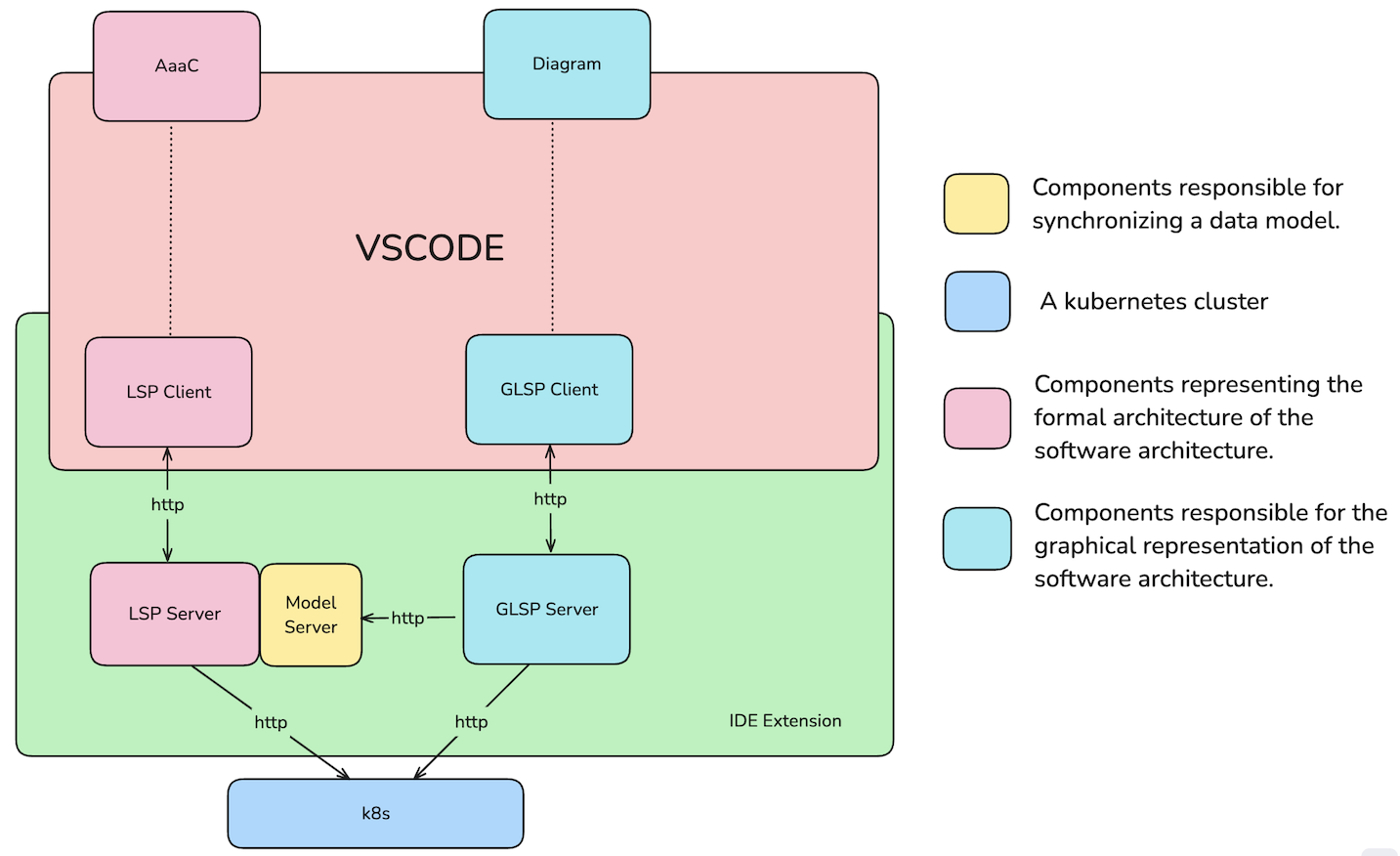}
    \caption{Architecture of Archer and its runtime-informed conformance-checking workflow}
    \Description{A component architecture diagram of Archer showing textual KDL editing, graphical editing, model synchronization, Kubernetes integration, recovery, and conformance diagnostics.}
    \label{fig:kdl-architecture}
\end{figure*}

Archer instantiates the proposed live architecture model workflow as a VS Code extension for Kubernetes deployment architecture. As shown in Fig.~\ref{fig:kdl-architecture}, the mechanism connects four paths: textual KDL editing, graphical architecture editing, shared model synchronization, and Kubernetes API integration for recovery and conformance diagnostics.

The architecture is based on the coordination of server-side components that implement language services, diagram services, view synchronization, and Kubernetes integration. The LSP server processes textual KDL documents and rebuilds their abstract syntax tree (AST) after edits. The GLSP server and client render and edit the graphical view. The Model Server mediates between these services: text edits are translated into diagram updates, and graphical edits are reflected back into the shared model. LSP, GLSP, Langium, and the Model Server are implementation mechanisms used to realize the live architecture model workflow; they are not claimed as research contributions by themselves.

One of the key functional features of the developed tool is its ability to extract the architectural structure directly from a running Kubernetes cluster. The current implementation separates two cluster-interaction paths. During conformance checking, Archer queries the Kubernetes API and reports diagnostics without modifying the model or the cluster. During recovery, Archer performs snapshot-based extraction: it queries the Kubernetes API, analyzes the deployed objects within the supported KDL scope, and inserts recovered KDL elements into the currently opened model. It does not implement an incremental runtime-update mechanism based on Kubernetes watch events or retained runtime snapshots.

\subsection{Modeling language}
\label{sec:modeling-language}
Archer represents namespaces containing pods, services, and ingresses, with nested containers, ports, and supported volume references. Service-to-pod and ingress-to-service links are inferred by the rules below. Appendix~\ref{app:representation} gives the metamodel and its containment structure.

\subsubsection{Kubernetes-to-KDL mapping semantics}
Recovery follows these field-level and edge-inference rules:

\begin{itemize}
    \item \textbf{Namespace mapping.} A \texttt{NamespaceNode} is created from each namespace name, excluding namespaces whose names start with \texttt{kube}.
    \item \textbf{Pod mapping.} \texttt{V1Pod} objects are mapped to \texttt{PodNode} objects with recovered containers, ports, and supported volume references. Controller kind is represented as a pod attribute when owner-reference resolution is available (e.g., Deployment/ReplicaSet and StatefulSet in the evaluated artifacts); resolution follows Pod--ReplicaSet--Deployment and direct Pod--StatefulSet ownership. Cardinality is read from the resolved controller's \texttt{spec.replicas}. The fallback label \texttt{RC} denotes ReplicationController in the notation but does not establish that such a controller exists when ownership is unresolved. For deployment-level views, the evaluated implementation keeps one representative pod per Deployment.
    \item \textbf{Service mapping.} \texttt{V1Service} objects are mapped to \texttt{ServiceNode} with \texttt{type}, \texttt{ports}, and inferred links to pod ports.
    \item \textbf{Ingress mapping.} \texttt{V1Ingress} objects are mapped at rule granularity (one \texttt{IngressNode} per rule), with links inferred from backend service name and backend service port.
    \item \textbf{Edge Rule 1 (Service $\rightarrow$ Pod Port).} For a service selector, matched pods are resolved by label conjunction (AND over selector keys). A service port is linked to a pod port when \texttt{targetPort} matches either pod-port number or pod-port name.
    \item \textbf{Edge Rule 2 (Ingress $\rightarrow$ Service Port).} For each ingress HTTP path, backend service and port are resolved; a link is created when backend port matches service-port number or name.
    \item \textbf{Port and volume mapping.} Pod/service ports are normalized to \texttt{PortNode(name, number)}. Volumes are recovered from \texttt{spec.volumes}, \texttt{env}, and \texttt{envFrom}; volume entities are deduplicated by \texttt{(name,type)}.
\end{itemize}

Within a recovered snapshot, mapping is executed in a fixed order per namespace: pods (including containers/ports/volumes), then services (including service-to-pod links), then ingresses (including ingress-to-service links). This order processes candidate target nodes before link inference; a link is created only when the matching rules resolve a supported target.

\subsubsection{Recovery workflow and model handling}
The \texttt{Recover} action is a model-modifying operation in the current prototype. Its input is the current Kubernetes API snapshot, and its output is a set of KDL elements inserted into the currently opened KDL document. Archer appends recovered namespaces, pods, services, ingresses, ports, volumes, and inferred links to the current semantic model and then serializes the modified model back to the textual KDL document. The current implementation does not perform identity-based merge with existing model elements, does not delete model elements that are absent from the cluster, and does not resolve conflicts between user-authored and recovered elements. If recovery is executed on a non-empty model that already contains resources with the same Kubernetes names, duplicate model elements can be introduced. For this reason, the current recovery workflow is intended primarily for initializing a KDL model from an empty or disposable document, or for producing a reference model for comparison.

Diagram information is generated for recovered elements. When no explicit location is supplied, newly recovered nodes receive default coordinates and dimensions, and recovered edges receive no routing points. Existing diagram attributes in the opened model are not used to compute an identity-based layout merge for recovered elements. Manual layout adjustment or a layout command may therefore be needed after recovery.

Conformance checking is separate from recovery. Validators compare the current KDL model with the live Kubernetes state and report mismatches as diagnostics. They do not modify the cluster, rewrite the model, or repair inconsistencies automatically.

\subsubsection{Coverage boundaries}
The original KDL notation was introduced as a compact graphical notation for describing architecturally relevant Kubernetes deployment objects rather than as a full YAML or API mirror \citep{KDLNotationDescribe}. Archer follows this architectural interpretation. Kubernetes objects form a much broader API surface: objects encode desired state and current state through \texttt{spec} and \texttt{status} fields \citep{kubernetesObjectsDocs}, workload APIs include controllers such as Deployments, StatefulSets, DaemonSets, Jobs, and CronJobs \citep{kubernetesWorkloadControllersDocs}, and separate API areas cover networking, storage, authorization, and extension mechanisms \citep{kubernetesIngressDocs,kubernetesVolumesDocs,kubernetesRBACDocs,kubernetesCustomResourcesDocs}. Table~\ref{tab:kdl-coverage-boundaries} therefore makes explicit which Kubernetes concepts are represented by the current Archer KDL metamodel and how the boundary affects recovery and checking.

\begin{table*}[htbp]
\centering
\scriptsize
\caption{Current KDL coverage boundary and practical effect on recovery and conformance checking}
\label{tab:kdl-coverage-boundaries}
\begin{tabular}{|p{0.16\textwidth}|p{0.18\textwidth}|p{0.25\textwidth}|p{0.29\textwidth}|}
\hline
\textbf{Kubernetes concept} & \textbf{KDL representation} & \textbf{Recovery support} & \textbf{Effect on conformance checking} \\
\hline
Namespaces & First-class \texttt{NamespaceNode}. & Supported for namespaces returned by the Kubernetes API, excluding namespaces whose names start with \texttt{kube}. & Namespace presence/absence can be checked. This prefix filter also excludes user namespaces such as \texttt{kube-demo}; it does not implement a complete system-namespace classification. \\
\hline
Pods, containers, and pod ports & First-class \texttt{PodNode}, \texttt{ContainerNode}, and \texttt{PortNode}. & Partially supported. Archer recovers selected pods and their containers/ports from the Kubernetes API; the evaluated implementation collapses deployment-level views to one representative pod. & Presence, container names, and port names/numbers can be checked for represented pods. Pod-template fields such as probes, resources, security context, scheduling rules, and labels are not represented or checked. \\
\hline
Workload controllers & \texttt{PodController} and \texttt{PodCardinality} attributes under a pod; controllers are not first-class model elements. & Partially supported. Controller kind and replica count can be represented for recovered pods when owner-reference resolution is available. Deployment, ReplicaSet, and StatefulSet values are represented as pod attributes; DaemonSet, Job, and CronJob are not recovered as first-class controller resources. & Controller kind and cardinality can be compared only as pod attributes. Controller-level architecture, rollout strategy, selectors, templates, job completion semantics, and schedules are outside the current checking scope. \\
\hline
Services and service ports & First-class \texttt{ServiceNode}, \texttt{ServiceTypeNode}, \texttt{PortNode}. & Supported for service name, type, and declared service ports. & Service presence, type, and port names/numbers can be checked. Headless-service semantics (\texttt{clusterIP: None}), \texttt{ExternalName}, EndpointSlice state, traffic policies, session affinity, and annotations are not comparison facts. \\
\hline
Service-to-pod links & References from \texttt{ServiceNode} to target \texttt{PortNode}. & Partially supported. Links are inferred from service selectors and target-port resolution when target pods and ports are represented. & Link checking is limited to represented service-to-pod relationships. Complex selector behavior, missing labels, non-pod backends, and EndpointSlice-only services can lead to missing links or unchecked relationships. \\
\hline
Ingress rules and ingress-to-service links & First-class \texttt{IngressNode} with \texttt{host}; links to service ports. & Partially supported. Archer recovers ingress rules and backend service links when the backend service/port is resolvable. & Ingress name and host can be checked. Path rules, \texttt{pathType}, TLS, ingress class, controller-specific annotations, and Gateway API resources are outside the current scope. \\
\hline
ConfigMaps and Secrets & Represented only as \texttt{VolumeNode} values of type \texttt{configmap} or \texttt{secret}. & Partially supported as references from pod volumes, \texttt{env}, and \texttt{envFrom}; the ConfigMap/Secret objects and key-level data are not recovered as first-class elements. & Presence/type of represented references can be checked. Object existence, keys, values, optional flags, and update semantics are not checked. \\
\hline
Other volume types & Generic \texttt{VolumeNode} syntax exists, but current recovery classifies only \texttt{secret} and \texttt{configmap}. & Unsupported for \texttt{persistentVolumeClaim}, \texttt{emptyDir}, \texttt{hostPath}, projected volumes, CSI volumes, and other volume kinds. & These volume kinds can cause false negatives: if they are architecturally relevant, Archer cannot fully recover or compare them with the current metamodel and extraction rules. \\
\hline
NetworkPolicy, RBAC, CRDs, and custom resources & No first-class KDL elements. & Unsupported. Archer does not query or recover these resources. & Policies, permissions, custom operators, domain-specific resources, and their drift are invisible to current recovery and validators. They require future metamodel extensions. \\
\hline
Helm and Kustomize inputs & No chart, values, overlay, or template representation. & Unsupported as authoring sources. Archer works from the live cluster state and KDL documents, not from Helm chart or Kustomize overlay structure. & Generated resources can be compared after deployment only if they fall inside the supported KDL resource subset. Chart values, overlays, and pre-deployment intent are not checked. \\
\hline
\end{tabular}
\end{table*}

The authoring, recovery, and comparison scopes are distinct: generic KDL volume nodes can record storage intent that the current extractor does not recover, while other Kubernetes concepts have no KDL representation. Accordingly, Archer currently checks an architectural subset of Kubernetes deployment structure. Authoring-scope facts included in the ground truth become false negatives when not recovered; unsupported concepts absent from the ground truth and perturbation set are invisible to the reported metrics.

\subsubsection{Helm and Kustomize workflows}
Many Kubernetes deployments are authored through Helm charts or Kustomize overlays rather than hand-written raw manifests. Helm supports local rendering of chart templates through \texttt{helm template} \citep{helmTemplateDocs}. Kubernetes also supports declarative object management through Kustomize, and \texttt{kubectl kustomize} builds a set of resources from a directory or URL containing a \texttt{kustomization.yaml} file \citep{kubernetesKustomizeDocs,kubectlKustomizeDocs}.

Archer reads live Kubernetes objects after deployment; it does not ingest charts, overlays, or rendered manifests. Rendered files can inform manual KDL authoring, but diagnostics cover only supported deployed facts. Changes confined to chart/overlay sources are outside the checking scope, and model elements cannot currently be traced back to the values, templates, or patches that produced them.

\subsection{Worked example}

\begin{figure*}[t]
    \centering
    \includegraphics[width=\textwidth]{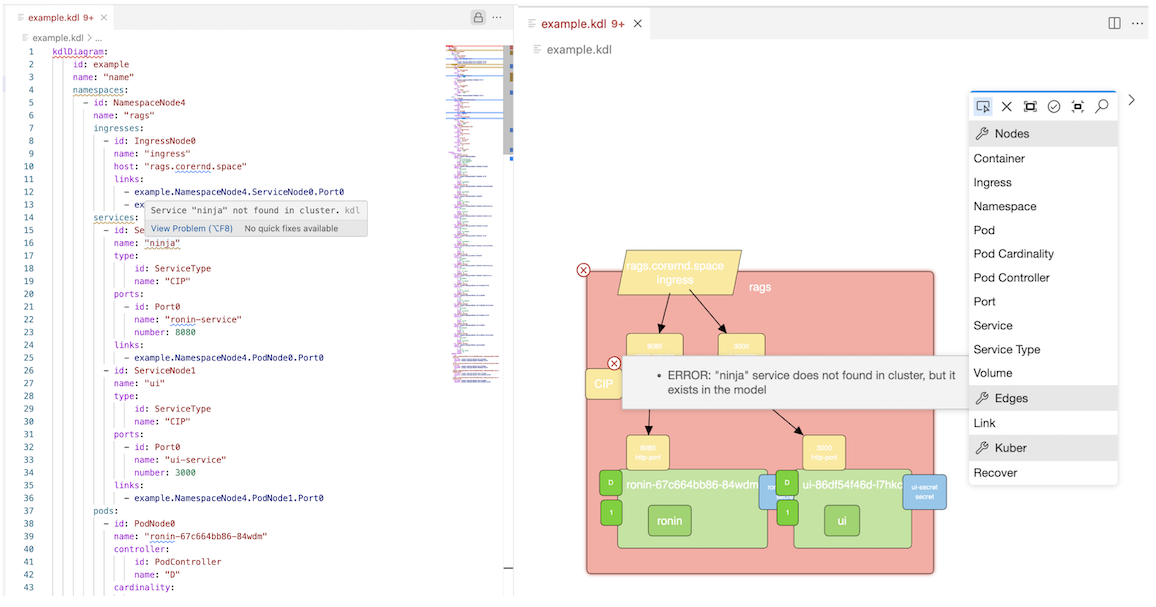}
    \caption{Archer interface showing textual KDL, synchronized graphical architecture, and conformance diagnostics}
    \Description{A screenshot of Archer showing the textual KDL model on the left, the synchronized graphical architecture on the right, and diagnostic messages for model-cluster mismatches.}
    \label{fig:kdl-example}
\end{figure*}

Figure~\ref{fig:kdl-example} shows the user workspace for designing a Kubernetes cluster architecture using the developed VS Code extension. The following worked example makes the same workflow explicit using the \texttt{model-serving-tensorflow} (abbreviated \texttt{model-serving-tf}) evaluation subject. The deployed application contains one namespace, one TensorFlow Serving pod behind a service, and one ingress route.

Starting from an empty document, recovery produces the supported namespace, ingress, service, pod, ports, and links. The saved reference and recovered KDL-schema excerpts are provided in Appendix~\ref{app:example}. Two differences illustrate the comparison boundary:
\begin{itemize}
    \item The reference ingress stores the path-regex \texttt{/tf(/|\$)(.*)} in the legacy \texttt{host} field; recovery supplies Archer's constructor default \texttt{localhost}. This is a representation/default mismatch under the scoring policy, not evidence of runtime drift. Hosts, paths, and defaults require separate treatment.
    \item The reference includes \texttt{persistentVolumeClaim} and \texttt{volumeMounts} entries that recovery omits, producing false negatives. Here, \texttt{volumeMounts} is a legacy generic-volume encoding of mount intent, not a Kubernetes volume kind. Generated \texttt{Pod.name} suffixes are excluded from attribute scoring; controller/cardinality are compared instead.
\end{itemize}

Figure~\ref{fig:kdl-example} shows the textual model and synchronized diagram with conformance diagnostics. After inspecting a diagnostic, the user can update KDL intent or source manifests, let an external GitOps controller reconcile the deployment, or classify the difference as intentionally outside the modeled scope.

\section{Evaluation}
The evaluation is an initial, scope-bounded assessment of two implemented Archer capabilities: (1) recovery of an architectural model from a Kubernetes cluster and (2) detection of inconsistencies between the architectural model and the current cluster state. The goal is to measure precision and recall for these capabilities within the supported KDL scope. The evaluation does not claim to establish industrial scalability, developer usefulness, or representativeness for large production Kubernetes deployments.

Three Kubernetes projects from the official Kubernetes examples repository\footnote{\url{https://github.com/kubernetes/examples}} were selected as worked examples. The subjects were selected by feature coverage rather than by representativeness: together, they exercise stateless deployments and services, stateful workloads, persistent-storage declarations, and label-based service-to-pod links. The repository was used because it provides public, stable, and reproducible manifests that readers can inspect at a fixed revision. These subjects are tutorial-scale examples rather than representative industrial deployments, so the results should be interpreted as early evidence for the implemented recovery and comparison rules.

\begin{itemize}
    \item model-serving-tf — a TensorFlow-based project demonstrating the deployment of a machine learning model using Deployment, Service, and Volume for storing model weights.

    \item Cassandra — a distributed database deployed via StatefulSet, Headless Service, and PersistentVolumeClaim, used to exercise represented stateful-controller attributes and expose storage-related coverage boundaries.

    \item Guestbook — a multi-component application including frontend, redis-master, and redis-slave, using Deployment, Service, and labels to link components. This example exercises service selectors and links among several deployment components.
\end{itemize}

Table~\ref{tab:subject-coverage} summarizes the selection rationale. The table also makes explicit where a subject includes Kubernetes constructs that are outside the current KDL coverage. In particular, Cassandra includes PersistentVolumeClaim resources, which are useful for exposing the current boundary of volume recovery but are not fully modeled by the current KDL implementation.

\begin{table}[htbp]
\centering
\caption{Evaluation subject selection rationale}
\label{tab:subject-coverage}
\small
\setlength{\tabcolsep}{4pt}
\begin{tabularx}{\linewidth}{|>{\raggedright\arraybackslash}p{0.23\linewidth}|>{\raggedright\arraybackslash}X|>{\raggedright\arraybackslash}X|}
\hline
\textbf{Subject} & \textbf{Covered constructs} & \textbf{Reason for inclusion} \\
\hline
model-serving-tf & Deployment, Service, container ports, model-storage volume declaration & Exercises recovery of a compact stateless deployment with service exposure and volume-related attributes. \\
\hline
Cassandra & StatefulSet, Headless Service, PersistentVolumeClaim declaration & Exercises represented StatefulSet controller/cardinality attributes and exposes the current limitation around persistent-storage modeling. \\
\hline
Guestbook & Multiple Deployments, Services, labels/selectors, service-to-pod links & Exercises recovery and comparison of several linked deployment components connected through service selectors. \\
\hline
\end{tabularx}
\end{table}

\subsection{Evaluation protocol and reproducibility}
\label{sec:evaluation-protocol}
To make the recovery results reproducible, we used paired artifacts per project: (i) a manually constructed ground-truth (GT) KDL model and (ii) a recovered KDL model produced by Archer from the running cluster state in an empty KDL document. GT models were created by inspecting Kubernetes manifests and the live cluster, then normalized to the KDL schema with deterministic per-type identifiers inside each namespace.

Scoring is deterministic and based on exact set matching in three dimensions (entities, links, attributes). Entities and links are matched using hierarchical path identifiers. Attribute comparison uses tuples of \texttt{(entity\_path, attribute\_name, attribute\_value)} under a predefined comparison policy. The recovery scoring pipeline computes TP/FP/FN, then precision and recall. Both result tables display ratios to four decimal places from the shown counts; the decimal format does not express statistical certainty. The denominators are small, including a single recovered link for Cassandra.

The public artifact is available at \url{https://github.com/AsakoKabe/archer-kdl}; the source and saved recovery data cited here are pinned to revision \href{https://github.com/AsakoKabe/archer-kdl/tree/70ff4d7e7f2f39df60e9a4905064d4e749d546ef}{\texttt{70ff4d7}}. It includes source code, build instructions, \texttt{demo.mp4}, and the \texttt{evaluate/} directory with \texttt{compute\_metrics.py}, dependencies, paired GT/recovered YAML models, and a manual detection protocol. Table~\ref{tab:eval-recovery} can be re-scored from the saved pairs; this does not repeat cluster extraction. Repeating Table~\ref{tab:eval-diff} requires a Kubernetes cluster, \texttt{kubectl}, VS Code, and manual execution and interpretation of diagnostics. The artifact provides the checklist and aggregate results, but not an automated detection harness or a complete trial-level diagnostic log.

The artifact README identifies the input manifests in the Kubernetes examples repository at commit \texttt{d6b8cd27}, under \path{AI/model-serving-tensorflow/}, \path{databases/cassandra/}, and \path{web/guestbook/all-in-one/}. It records a local environment with macOS 15.7.3 (arm64), Docker Desktop 4.24.0, Docker Engine 24.0.6, Minikube 1.32.0, Kubernetes profile 1.28.3, and \texttt{kubectl} 1.28.4. Exact historical VS Code, Langium, and GLSP runtime versions are not recorded in that environment table.

\subsection{Recovery of architectural model from the cluster}

Starting from an empty KDL document, the Archer tool recovered a snapshot architectural model from data obtained from the Kubernetes API. The recovered model was compared with a reference model manually created based on official documentation and YAML files.

Recovery is scored separately for entities, attribute tuples, and links. In each dimension, the following definitions apply:

\begin{itemize}
    \item True Positive (TP) — the number of facts present in both the recovered and reference sets.
    \item False Positive (FP) — the number of facts present only in the recovered set.
    \item False Negative (FN) — the number of facts present only in the reference set.
    \item Precision — the proportion of recovered facts that match the reference:
    \begin{equation} \label{eq:precision}
    Precision = \frac{TP}{TP + FP}
    \end{equation}
    \item Recall — the proportion of reference facts recovered:
    \begin{equation} \label{eq:recall}
    Recall = \frac{TP}{TP + FN}
    \end{equation}
\end{itemize}

Attribute names are matched exactly after case-preserving canonicalization, and values are compared after schema-aware normalization. The comparison uses a fixed attribute subset per entity type (e.g., for pods, \texttt{controller} and \texttt{cardinality} are compared, while \texttt{Pod.name} is excluded to avoid false mismatches caused by runtime pod-name suffixes). Scalar formatting and whitespace are normalized, and recovery defaults are applied consistently for optional fields (e.g., \texttt{Ingress.host=localhost}, \texttt{Port.number=8080}, \texttt{Port.name=Port\{index\}}, and \texttt{PodController=RC} when no owner is resolved). Defaults produced by this policy, when absent in the reference model, are counted as \textit{FP}; conversely, reference attributes not recovered are counted as \textit{FN}.

\begin{table*}[htbp]
\centering
\small
\setlength{\tabcolsep}{3pt}
\caption{Recovery scoring over saved GT/recovered model pairs using \texttt{compute\_metrics.py}. Counts denote model entities, attribute tuples, and links; ratios use the shown counts.}
\begin{tabular}{|l|ccccc|ccccc|ccccc|}
\hline
\textbf{Project}
& \multicolumn{5}{c|}{\textbf{Entities}} 
& \multicolumn{5}{c|}{\textbf{Attributes}} 
& \multicolumn{5}{c|}{\textbf{Links}} \\
\cline{2-16}
& TP & FP & FN & Precision & Recall 
& TP & FP & FN & Precision & Recall 
& TP & FP & FN & Precision & Recall \\
\hline
model-serving-tf 
& 9 & 0 & 2 & 1.0000 & 0.8182
& 15 & 1 & 5 & 0.9375 & 0.7500
& 3 & 0 & 0 & 1.0000 & 1.0000 \\
\hline
cassandra 
& 9 & 0 & 1 & 1.0000 & 0.9000
& 16 & 0 & 2 & 1.0000 & 0.8889
& 1 & 0 & 0 & 1.0000 & 1.0000 \\
\hline
guestbook 
& 16 & 0 & 0 & 1.0000 & 1.0000
& 28 & 0 & 0 & 1.0000 & 1.0000
& 3 & 0 & 0 & 1.0000 & 1.0000 \\
\hline
\end{tabular}
\label{tab:eval-recovery}
\end{table*}

\begin{table*}[htbp]
\centering
\small
\setlength{\tabcolsep}{3pt}
\caption{Inconsistency detection under the manual perturbation protocol. Counts denote diagnostic outcomes, not recovered model elements; ratios use the shown counts.}
\begin{tabular}{|l|ccccc|ccccc|ccccc|}
\hline
\textbf{Project}
& \multicolumn{5}{c|}{\textbf{Entities}} 
& \multicolumn{5}{c|}{\textbf{Attributes}} 
& \multicolumn{5}{c|}{\textbf{Links}} \\
\cline{2-16}
& TP & FP & FN & Precision & Recall 
& TP & FP & FN & Precision & Recall 
& TP & FP & FN & Precision & Recall \\
\hline
model-serving-tf 
& 12 & 0 & 2 & 1.0000 & 0.8571
& 6 & 1 & 1 & 0.8571 & 0.8571
& 3 & 0 & 0 & 1.0000 & 1.0000 \\
\hline
cassandra 
& 8 & 0 & 1 & 1.0000 & 0.8889
& 6 & 0 & 1 & 1.0000 & 0.8571
& 3 & 0 & 0 & 1.0000 & 1.0000 \\
\hline
guestbook 
& 8 & 0 & 0 & 1.0000 & 1.0000
& 6 & 0 & 0 & 1.0000 & 1.0000
& 3 & 0 & 0 & 1.0000 & 1.0000 \\
\hline
\end{tabular}
\label{tab:eval-diff}
\end{table*}

Table~\ref{tab:eval-recovery} reports agreement with the saved reference models on the three selected examples. The remaining FP and FN cases are analyzed after the detection protocol because the same coverage boundary affects both recovery and checking.

\subsection{Detection of inconsistencies between the model and the cluster}

Table~\ref{tab:eval-diff} evaluates inconsistency detection in a manual protocol that requires a running Kubernetes cluster and interactive use of Archer in VS Code. Unlike Table~\ref{tab:eval-recovery}, each trial is executed as a controlled perturbation and checked against diagnostics produced by Archer validators. The perturbations are author-defined and coverage-based: they are intended to exercise the mismatch classes represented in the current KDL comparison model, not to approximate a random sample of production drift events. This ties the protocol to explicit model semantics, but the author-defined selection can yield optimistic performance; a documented manual protocol does not by itself provide a complete audit trail.

Table~\ref{tab:perturbation-taxonomy} defines the perturbation taxonomy used in the inconsistency-detection evaluation. The taxonomy is derived from the three comparison dimensions used throughout the evaluation: entities, links, and attributes. Entity perturbations test whether the model and cluster contain the same architectural elements. Link perturbations test whether recovered service-to-pod and ingress-to-service relationships match the intended relationships. Attribute perturbations test whether selected properties of matched elements have matching values. Model-only denotes edits made in KDL, including references to absent cluster objects and values contradicting cluster state. The edit side does not by itself distinguish intra-model well-formedness from model--cluster conformance. Results are aggregated by fact dimension; no separate Model-only breakdown is available.

\begin{table}[htbp]
\centering
\caption{Perturbation taxonomy for inconsistency detection}
\label{tab:perturbation-taxonomy}
\begin{tabular}{|l|p{0.39\linewidth}|p{0.25\linewidth}|}
\hline
\textbf{Class} & \textbf{Examples} & \textbf{Purpose} \\
\hline
Entity & Add or remove pod, service, ingress, or volume & Element presence mismatch \\
\hline
Link & Break or add service selector and KDL link references & Relationship mismatch \\
\hline
Attribute & Change ingress host, service type, pod controller, or pod cardinality & Property mismatch \\
\hline
Model-only & Reference absent cluster entities or contradict cluster attribute values & Model-side inconsistency \\
\hline
\end{tabular}
\end{table}

For each project, we execute the following procedure:
\begin{enumerate}
    \item Deploy the project to a Kubernetes cluster.
    \item Recover the architecture model with Archer (or open an existing KDL model).
    \item Apply one controlled modification from one side only: cluster side (A) or model side (B).
    \item Trigger Archer's conformance check.
    \item Record diagnostics (errors/warnings) in VS Code.
    \item Compare diagnostics with the known injected inconsistency and count TP/FP/FN.
    \item Revert the modification before the next trial.
\end{enumerate}

Each trial introduces exactly one perturbation. Perturbations are grouped as follows:
\begin{itemize}
    \item \textit{Entity perturbations (add/remove):} \texttt{pod}, \texttt{service}, \texttt{ingress}, and \texttt{volume}. Side A edits cluster resources via \texttt{kubectl} (or deployment volume mounts for volumes), while side B edits corresponding KDL nodes.
    \item \textit{Link perturbations:} break/add links by changing service selectors on side A, or by removing/adding link references in KDL on side B.
    \item \textit{Attribute perturbations:} \texttt{ingress.host}, \texttt{service.type}, \texttt{pod.controller}, and \texttt{pod.cardinality}, changed in cluster objects (A) or in KDL (B).
    \item \textit{Model-only perturbations:} references to non-existent entities and attribute values intentionally contradicting the cluster state.
\end{itemize}

For each project and each dimension (entities, attributes, links), we compare two sets: \textit{GT} (known injected inconsistencies) and \textit{Pred} (diagnosed inconsistencies). We compute \textit{TP} $=|Pred \cap GT|$, \textit{FP} $=|Pred \setminus GT|$, and \textit{FN} $=|GT \setminus Pred|$, then report precision and recall using Equations~\eqref{eq:precision} and~\eqref{eq:recall}.

The recorded operation summary lists entity/attribute/link operation instances of 14/7/3 for \texttt{model-serving-tf}, 9/7/3 for Cassandra, and 8/6/3 for Guestbook. The corresponding expected diagnostic-outcome counts have the same values, totaling 24, 19, and 17. These are aggregate records, not a complete per-operation log or evidence of independent trials. Recall uses TP+FN and precision uses TP+FP separately within each dimension; the totals are not a common denominator for both metrics. The record does not provide a separate count by edit side or Model-only class.

Table~\ref{tab:eval-diff} reports detection outcomes for the controlled, author-defined perturbations. These results concern the tested mismatch classes and do not establish detection of arbitrary production drift.

\subsection{Error analysis}
Tables~\ref{tab:eval-recovery} and~\ref{tab:eval-diff} must be interpreted together with the coverage boundary in Table~\ref{tab:kdl-coverage-boundaries} and the error analysis in Table~\ref{tab:error-analysis}. Archer compares only Kubernetes concepts represented in the current KDL metamodel and recovered by the current implementation. Unsupported resources affect the metrics in two different ways. If an unsupported concept is included in the ground-truth KDL model or in a perturbation, it becomes an FN because Archer cannot recover or diagnose it. If the same concept is absent from the ground truth and from the perturbation set, it is invisible to the metric and therefore does not reduce the aggregate score. The aggregate scores therefore depend on both the comparison scope and the selected test facts.

\begin{table*}[htbp]
\centering
\scriptsize
\caption{Root causes of FP and FN in recovery and inconsistency detection}
\label{tab:error-analysis}
\setlength{\tabcolsep}{3.5pt}
\begin{tabular}{|>{\raggedright\arraybackslash}p{0.17\textwidth}|>{\raggedright\arraybackslash}p{0.17\textwidth}|>{\raggedright\arraybackslash}p{0.28\textwidth}|>{\raggedright\arraybackslash}p{0.30\textwidth}|}
\hline
\textbf{Root cause} & \textbf{Category} & \textbf{Metric effect} & \textbf{Observed example} \\
\hline
Unsupported persistent-storage concepts & Generic storage encoding and recovery gap & Missing volume entities and their attributes are counted as FN in recovery. Detection can also miss volume-related perturbations because the compared model does not contain a recoverable runtime counterpart. & In \texttt{model-serving-tf}, the ground-truth model contains \texttt{persistentVolumeClaim} and \texttt{volumeMounts} entries for \texttt{model-volume}; the recovered model omits them. In \texttt{cassandra}, the ground truth contains a \texttt{volumeMounts} entry for \texttt{cassandra-data}; the recovered model omits it. \\
\hline
Narrow volume extraction & Recovery implementation limitation & Even though KDL has generic \texttt{VolumeNode} syntax, the current extractor recovers only selected configuration references such as \texttt{secret} and \texttt{configmap}. Other volume kinds become FN when they appear in ground truth. & The missing \texttt{persistentVolumeClaim} and \texttt{volumeMounts} entries in the evaluated artifacts are not produced by the current recovery rules. \\
\hline
Ingress host/default mismatch & GT representation and Archer constructor default & A defaulted or normalized value can create an FP for the recovered/default attribute and an FN for the expected attribute value. & In \texttt{model-serving-tf}, the ground-truth KDL ingress \texttt{host} attribute stores \texttt{/tf(/|\$)(.*)}, while the recovered model contains \texttt{localhost}. This contributes to the attribute FP/FN around ingress attributes. \\
\hline
Runtime-generated names and canonicalization & Evaluation-policy decision & Some Kubernetes runtime values are intentionally not compared to avoid meaningless mismatches; this can make the metric less sensitive to name-level drift. & Runtime pod names include suffixes, e.g., \texttt{tf-serving-8474cc5bb7-fwdcq}, while the architectural model names the logical workload \texttt{tf-serving}. The scoring policy excludes \texttt{Pod.name} and compares controller/cardinality instead. \\
\hline
Author-defined perturbation set & Evaluation-design limitation & Precision and recall can be optimistic because tested inconsistencies are selected to exercise current validators. Unsupported Kubernetes resources, such as NetworkPolicy, RBAC, CRDs, Jobs, CronJobs, Helm values, and Kustomize overlays, are not represented in the perturbation space. & Table~\ref{tab:eval-diff} reports high detection scores for entity, link, and attribute perturbations covered by current validators, but does not test arbitrary production drift or independently sampled incidents. \\
\hline
\end{tabular}
\end{table*}

The saved recovery pairs give a specific error breakdown: \texttt{model-serving-tf} has two storage-entity FNs, four storage-attribute FNs, and one ingress-host attribute FN; the defaulted host also contributes one FP. Cassandra has one storage-entity FN and two storage-attribute FNs. Guestbook has no errors in the saved pair, consistent with coverage of this selected example rather than evidence of general reliability. Thus, most FNs reflect facts present in the authoring/GT scope but outside recovery, rather than an inability to express every such fact in generic KDL syntax. The dominant FP pattern is not spurious graph recovery but default-value mismatch: Archer currently materializes some optional attributes, and the evaluation counts those values when they differ from the reference model. Future improvements should therefore separate three work items: extend the KDL metamodel for currently unsupported architectural concepts, extend recovery rules for those concepts, and refine the comparison policy for Kubernetes defaults and generated runtime identifiers.

\section{Threats to Validity}

\textbf{Internal validity.} The inconsistency-detection scenarios were created by the authors and may therefore be aligned with the validators implemented in Archer. Although the perturbation checklist covers entity, link, and attribute mismatches from both the model side and the cluster side, it is still a controlled set of changes rather than an independently sampled set of real drift events. The ground-truth KDL models and the interpretation of VS Code diagnostics were also produced by the authors, which may introduce judgment bias. We mitigated this risk by using deterministic matching rules, explicit TP/FP/FN definitions, paired ground-truth and recovered artifacts, and a documented perturbation checklist, but these measures do not establish independence of the oracle or scenario selection. Independent replication remains necessary. No baseline comparison is reported, so the results do not establish superior accuracy over alternative tools or simpler comparators.

\textbf{External validity.} The evaluation uses three tutorial-scale applications from the official Kubernetes examples repository. These subjects are useful for exercising the current KDL mapping, but they are not representative of large production Kubernetes deployments. The evaluation does not cover multi-namespace systems, source-level Helm or Kustomize workflows, large clusters, production naming conventions, runtime failure modes, repeated evolution sequences, or the full Kubernetes API. The results should therefore be interpreted as initial evidence for the supported KDL scope, not as evidence of industrial scalability or broad Kubernetes coverage.

\textbf{Construct validity.} Precision and recall measure whether Archer recovers selected model elements and reports selected model-cluster inconsistencies under the stated comparison policy. These metrics do not measure scalability, conformance-check latency, recovery latency, view-synchronization correctness under complex edits, whether developers or architects find the synchronized textual and graphical workflow useful, whether the tool reduces maintenance effort, or whether the approach achieves direct erosion control. The current evidence supports recovery and detection claims within the evaluated scope; user-facing usefulness, scalability, robustness, and direct erosion-control effects require separate empirical studies.

\textbf{Reproducibility.} Recovery scores can be reproduced from the saved model pairs and scoring script at the public artifact revision identified in Section~\ref{sec:evaluation-protocol}. Re-extracting those models and repeating the manual detection protocol require the runtime environment and interactive use of Archer. Kubernetes versions, cluster configuration, extension state, and diagnostic interpretation can affect replication. The aggregate detection counts cannot be independently audited trial by trial from the available record; automated perturbation execution and diagnostic logging remain future work.

\section{Conclusion and Future Work}
This work presented live architecture models for cloud-native AaC, instantiated the idea through KDL and deterministic Kubernetes recovery and comparison rules, and realized it in Archer, a VS Code prototype for synchronized textual and graphical Kubernetes architecture modeling. Using KDL, users can create formalized descriptions of selected deployment infrastructure elements represented by the current metamodel.
A key feature of the tool is the extraction of selected architectural facts from a running Kubernetes cluster. This capability supports periodic and on-demand conformance checking between the design-time architectural description and the runtime cluster state. Feasibility checks on three feature-selected Kubernetes example applications under an author-defined protocol measured the precision and recall of architectural model recovery and inconsistency detection against the actual cluster state. The results provide initial evidence for the feasibility of the paper's thesis: a live architecture model can connect editable cloud-native architectural intent with selected runtime deployment facts and expose selected model-cluster inconsistencies within the supported KDL scope.

Archer is available at \url{https://github.com/AsakoKabe/archer-kdl}, with the pinned source and evaluation revision documented in Section~\ref{sec:evaluation-protocol}. Future work should extend resource coverage and comparison policies, investigate manifest generation and source traceability, and assess scale, runtime robustness, repeated evolution, view synchronization, and developer usefulness. Appendix~\ref{app:future-evaluation} details the proposed evaluation agenda; these outcomes have not yet been measured.

\bibliographystyle{ACM-Reference-Format}
\bibliography{sample-base}

\onecolumn
\appendix
\section{KDL representation and conceptual mapping}
\label{app:representation}
This appendix records representation details supplementary to Section~\ref{sec:modeling-language}. The conceptual mapping in Table~\ref{tab:dep-arch} is illustrative and does not establish compliance with an architectural-description standard.

\begin{table}[H]
\caption{Illustrative mapping of architecture-description concepts to Kubernetes artifacts}
\centering
\setlength{\tabcolsep}{4pt}
\begin{tabular}{|c|c|p{6.8cm}|}
\hline
\textbf{Conceptual Component} & \textbf{System Artifact} & \textbf{Description} \\ \hline

\textbf{Architectural Description (AD)} 
    & Kubernetes Deployment Architecture 
    & A collection of descriptions reflecting the structure and architectural principles of the system. Includes models, views, and perspectives. \\ \hline

\textbf{Model} 
    & KDL (Kubernetes Deployment Language) 
    & An architectural model represented as a domain-specific language (DSL) that expresses structural decisions at the configuration level. \\ \hline

\textbf{Source Code / Implementation} 
    & Kubernetes manifests: ingress, service, etc. 
    & YAML files that concretely implement the architectural decisions and system components. \\ \hline

\textbf{System} 
    & Kubernetes cluster 
    & The deployed system that realizes the architecture described in the model. \\ \hline

\textbf{Environment} 
    & Infrastructure context (e.g., node OS) 
    & The infrastructural context in which the Kubernetes cluster and the entire system operate. \\ \hline

\end{tabular}
\label{tab:dep-arch}
\end{table}

\begin{figure}[H]
    \centering
    \includegraphics[width=0.70\textwidth]{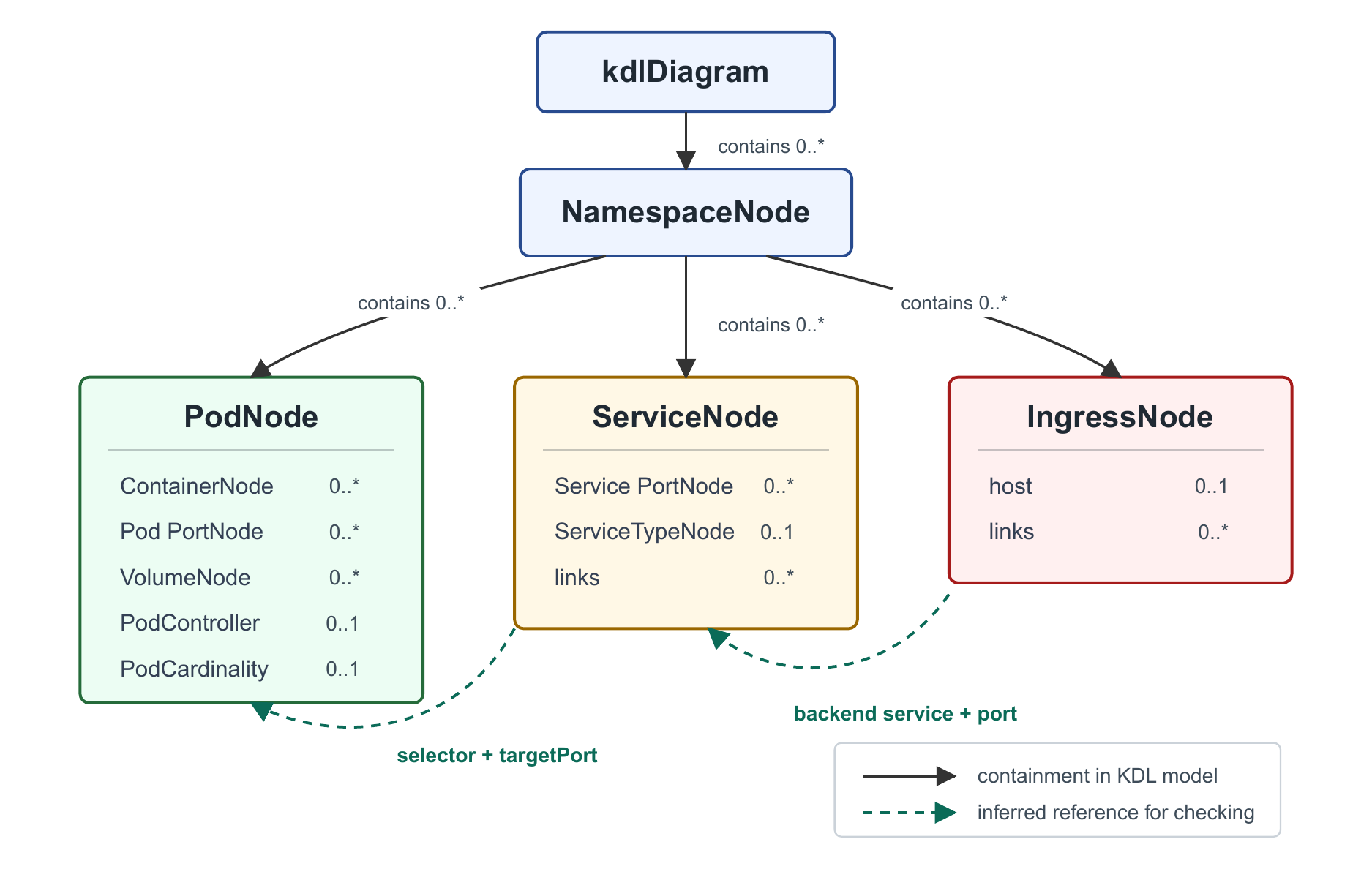}
    \caption{Supported Archer KDL metamodel for Kubernetes deployment architecture}
    \Description{A metamodel diagram showing kdlDiagram, NamespaceNode, PodNode, ServiceNode, and IngressNode. Solid arrows denote containment in the KDL model. Dashed arrows denote inferred references used for conformance checking from services to pods and from ingresses to services.}
    \label{fig:kdl-metamodel}
\end{figure}

The metamodel is organized as a hierarchical structure with \texttt{kdlDiagram} as the root. It contains \texttt{NamespaceNode} elements; each namespace contains \texttt{PodNode}, \texttt{ServiceNode}, and \texttt{IngressNode} elements. Pods contain \texttt{ContainerNode}, \texttt{PortNode}, \texttt{VolumeNode}, optional \texttt{PodController}, and optional \texttt{PodCardinality}. Services contain \texttt{ServiceTypeNode}, \texttt{PortNode}, and \texttt{links}; ingresses contain \texttt{host} and \texttt{links}. Solid arrows in Fig.~\ref{fig:kdl-metamodel} denote containment in the KDL model. Dashed arrows denote inferred references used by recovery and conformance checking: services link to pods through selector and \texttt{targetPort} matching, while ingresses link to services through backend service and port resolution. Every metamodel element has mandatory \texttt{id} and \texttt{name} fields.

In addition to the logical structure, the metamodel supports the description of graphical characteristics through the \texttt{diagram} section. Each model element can specify visual parameters, such as:

\begin{itemize}
    \item diagram coordinates (\texttt{x}, \texttt{y});
    \item dimensions (\texttt{width}, \texttt{height});
    \item \texttt{edgeAttributes}, which define the source and target vertices of the graph.
\end{itemize}

\newpage
\section{Worked-example model excerpts}
\label{app:example}
These abridged excerpts use the normalized YAML representation of KDL-schema elements in the saved evaluation artifacts, not Kubernetes manifests. They retain the reference/default and storage differences discussed in the main text; unrelated fields are omitted.

\noindent\begin{minipage}[t]{0.49\textwidth}
\subsection{Reference model}
\begingroup
\footnotesize
\begin{verbatim}
kdlDiagram:
  id: tf-serving-example
  namespaces:
    - id: NamespaceNode0
      name: "model-serving"
      ingresses:
        - id: IngressNode0
          name: "tf-serving-ingress"
          host: "/tf(/|$)(.*)"
          links:
            - tf-serving-example.NamespaceNode0.ServiceNode0.Port1
      pods:
        - id: PodNode0
          name: "tf-serving"
          volumes:
            - id: Volume0
              name: "model-volume"
              type: "persistentVolumeClaim"
            - id: Volume1
              name: "model-volume"
              type: "volumeMounts"
\end{verbatim}
\endgroup

\end{minipage}\hfill\begin{minipage}[t]{0.49\textwidth}
\subsection{Recovered model}
\begingroup
\footnotesize
\begin{verbatim}
kdlDiagram:
  id: tf-serving-example
  namespaces:
    - id: NamespaceNode0
      name: "model-serving"
      ingresses:
        - id: IngressNode0
          name: "tf-serving-ingress"
          host: "localhost"
          links:
            - tf-serving-example.NamespaceNode0.ServiceNode0.Port1
      pods:
        - id: PodNode0
          name: "tf-serving-8474cc5bb7-fwdcq"
          containers:
            - id: ContainerNode0
              name: "tensorflow-serving"
          ports:
            - id: Port0
              name: "PortNode0"
              number: 8500
            - id: Port1
              name: "PortNode1"
              number: 8501
\end{verbatim}
\endgroup

\end{minipage}

\section{Proposed evaluation agenda}
\label{app:future-evaluation}
\subsection{Developer and architect assessment}
A proposed follow-up study would recruit Kubernetes-experienced developers or architects for five tasks: author a KDL model, inspect its diagram, recover and compare a model, diagnose injected drift, and edit either view to check synchronization. Measures would include completion, time, modeling errors, diagnostic interpretation, unresolved inconsistencies, and qualitative feedback on usefulness and missing concepts. An initial walkthrough with two or three experts could identify confusing terminology and interactions before a larger controlled study; it would not establish statistical effectiveness. No participant study is reported here.

\subsection{Broader evaluation agenda}
Table~\ref{tab:broader-eval-agenda} outlines measurements needed to extend the present evidence.

\begin{table}[H]
\centering
\scriptsize
\caption{Planned evaluation dimensions beyond the current precision/recall study}
\label{tab:broader-eval-agenda}
\begin{tabular}{|p{0.18\textwidth}|p{0.25\textwidth}|p{0.25\textwidth}|p{0.22\textwidth}|}
\hline
\textbf{Dimension} & \textbf{Current coverage} & \textbf{Needed extension} & \textbf{Measurements} \\
\hline
Scale & Three tutorial-scale applications with small numbers of namespaces, pods, services, and ingresses. & Synthetic and real deployments with increasing numbers of namespaces, workload controllers, services, ingresses, ports, and links. & Recovery time, checking latency, memory use, model size, diagnostic count, and diagram update time. \\
\hline
Kubernetes heterogeneity & Selected deployment objects inside the current KDL scope. & Workloads and resources outside the current scope, including DaemonSets, Jobs, CronJobs, NetworkPolicies, RBAC resources, CRDs, custom operators, and richer storage/networking resources. & Per-resource recovery/checking coverage, FP/FN by resource kind, and unsupported-resource reporting. \\
\hline
Delivery workflows & Raw manifests from the official Kubernetes examples repository. & Helm, Kustomize, and GitOps-managed deployments, including rendered manifests and live clusters produced from chart/overlay sources. & Ability to recover deployed architectural structure, detect supported drift after deployment, and trace model elements to source artifacts where supported. \\
\hline
Runtime robustness & Controlled perturbations applied one at a time. & Failed deployments, partial rollouts, generated names, defaulted fields, deleted resources, renamed resources, and concurrent cluster changes. & Diagnostic stability, false positives from defaults, false negatives from missing resources, and behavior under repeated checks. \\
\hline
Evolution over time & Single recovery and single-step perturbations. & Multi-step model and cluster evolution, repeated recovery, and repeated conformance checks across several change sequences. & Duplicate model elements, stale diagnostics, recovery idempotence, model identity preservation, and accumulated FP/FN. \\
\hline
View synchronization & Demonstrated implementation of synchronized textual and graphical KDL editing. & Systematic edit sequences in text and diagram views, including invalid intermediate states and recovery after edit conflicts. & View-synchronization correctness, lost edits, layout preservation, parse/diagram consistency, and user-visible errors. \\
\hline
Developer/architect usefulness & Not evaluated in the current study. & Controlled tasks or expert walkthroughs as described above. & Task completion, time, errors, diagnostic interpretation accuracy, and qualitative usefulness feedback. \\
\hline
\end{tabular}
\end{table}

\end{document}